\documentclass[%
 reprint,showkeys,onecolumn,
 amsmath,amssymb,
 aps,
]{revtex4-2}
\usepackage{xcolor}
\usepackage{graphicx}
\usepackage{dcolumn}
\usepackage{bm}

\begin{document}

\preprint{APS/123-QED}

\title{{\AE}ther Coupling Effects on Casimir Energy for Self-Interacting Scalar Field within Extra Dimension}

\author{M. A. Valuyan}
\altaffiliation[Also at ]{Energy and Sustainable Development Research Center, Semnan Branch, Islamic Azad University, Semnan, Iran}
\email{m-valuyan@sbu.ac.ir;ma.valouyan@iau.ac.ir}
\affiliation{%
 Department of Physics, Semnan Branch, Islamic Azad University, Semnan, Iran.
}%




\date{\today}

\begin{abstract}
This paper presents comprehensive calculations for thermal and first-order radiative corrections to the Casimir energy in systems involving self-interacting massive and massless scalar fields coupled with {\ae}ther in a fifth compact dimension. The method used to compute the radiative correction to the Casimir energy differs from conventional approaches by applying a unique renormalization scheme that is consistent with specific boundary conditions or backgrounds.  Despite this divergence from conventional methodologies, our results demonstrate consistency within established physical limits. Furthermore, employing a toy model, we calculated the total Casimir energy density in the bulk, taking into account both thermal and radiative corrections. We also provide a thorough characterization of the total Casimir energy density in the compact dimension, detailing its magnitude and sign using graphical representations and quantitative data.

\end{abstract}

\keywords{Casimir Energy; {\AE}ther; Radiative Correction; Thermal Correction; Extra Dimension }
\maketitle

\section{\label{Intro.}Introduction}
Consider two uncharged conductive plates placed a few nanometers apart in a vacuum. According to classical physics, no field should exist between the plates due to the absence of an external field. However, quantum electrodynamic vacuum reveals a different picture. This vacuum is not truly empty; it is filled with transient electromagnetic waves. These virtual particles, or virtual photons, spontaneously appear and disappear. When the plates are placed in close proximity, they influence these virtual photons, resulting in a net force between the plates. This force, depending on the arrangement of the plates, can manifest as either attraction or repulsion. This phenomenon is known as the Casimir effect, named after Dutch physicist H. B. G. Casimir, who first theorized it in 1948 \cite{h.b.g.}. The Casimir effect remained an intriguing concept until 1958 when direct experimental measurements by Marcus Sparnaay confirmed its existence \cite{spaarnay,other.works.1}. Although these measurements did not have sufficient accuracy, they marked a pivotal moment in our understanding of this subtle force. In 1997, Lamoreaux measured the Casimir force using a torsion pendulum at distances ranging from 0.6 to 6 micrometers, finding good agreement with theoretical predictions \cite{Lamerux}. Subsequent irrefutable experimental confirmations of the Casimir force in the submicron regime, achieved through Atomic Force Microscopy (AFM) techniques, have had widespread implications across various fields. These precise measurements have not only reinforced the fundamental understanding of quantum field theory but have also propelled advancements in nanotechnology, chemistry, and biology \cite{Physica.scr.1,Physica.scr.2}. Since the Casimir energy assumes significance in systems with sizes smaller than a micrometer, its influence in the realm of biophysics has garnered more attention\cite{bio.Casimir.1}. In this regard, investigations into Casimir energy within biological systems, such as DNA and cell membrane proteins, have become a subject of study\cite{bio.Casimir.2,bio.Casimir.3}. The ability to manipulate and measure the Casimir force at such small scales has opened new avenues for innovation, from designing nanoscale devices to exploring molecular interactions. Beyond mere curiosity, investigations into Casimir energy extended beyond force calculations in multiple geometries for various quantum fields\cite{other.works.2}. Researchers delved into radiative and thermal corrections, exploring a wide range of fields and boundary conditions\cite{RC.1,JPhysG.Man}. The initial attempt to calculate radiative corrections to the Casimir energy of the electromagnetic field was conducted by Bordag et al.\cite{Bordag.1,Bordag.2,Bordag.3}. These corrections account for the effects of higher-order quantum electrodynamics interactions and have also been developed for other quantum fields, such as scalar fields and Dirac fermionic fields. Understanding radiative effects is crucial for applications in nanotechnology and materials science, where precise manipulation of forces at the nanoscale is essential. In addition to radiative corrections, thermal corrections significantly influence the behavior of the Casimir effect at finite temperatures. Unlike the idealized zero-temperature scenario, real-world applications often involve environments where thermal fluctuations cannot be ignored \cite{thermal.Brevik.}. Another aspect of considering the Casimir energy pertains to the mathematical methods developed to address inherent divergences and renormalize bare parameters within Lagrangian formulations \cite{all.method.ZO}. Notable regularization techniques include Zeta function regularization, the heat kernel series, the box subtraction scheme, and the Green's function formalism \cite{methods.ZetaF,Heat.kernel,cognola, methods.BSS}. These methods are crucial for obtaining finite and physically meaningful results from calculations involving the Casimir effect in quantum field theory. An important intriguing aspect of the Casimir effect involves the preservation or non-preservation of Lorentz symmetry and its effects on the values of Casimir energy\cite{cruz.1,cruz.2,cruz.Fermion}. As high energies disrupt Lorentz symmetry, researchers—often astronomers—have focused more attention on this topic and meticulously investigated the interplay of Casimir energy with various directions of Lorentz-violating quantum fields, yielding valuable results\cite{Lv.1,Aether.EM.,Mojavezi.}. An interesting idea is that the universe started in a very symmetric state where all spatial dimensions were compactified with equal radii, for example, topologically as a 4-dimensional torus. In general, the Casimir energies in this compact universe generate a stabilized potential for the radius of all directions. At very early times, however, the energy density of matter and radiation would be the dominant contribution, destabilizing the moduli fields and causing all directions to expand. If Lorentz symmetry is spontaneously broken in some direction, with a non-vanishing {\ae}ther field pointing in the fifth direction, it will slow down the dynamics of the moduli field associated with the broken direction. This broken direction will become compactified at a stabilized radius, while the unbroken directions are allowed to expand. Consequently, we might observe only the large three-dimensional space that preserves Lorentz symmetry, while the Lorentz-violating directions remain compactified. This cosmological scenario may establish a connection between the dimensionality of spacetime and the violation of Lorentz symmetry. Numerous works have discussed various aspects of this idea \cite{9ta12.0905.0328.1,9ta12.0905.0328.2,9ta12.0905.0328.3,9ta12.0905.0328.4}. One important issue in discussing this idea is the consideration of the Casimir effect, which has emerged as a significant factor in addressing some of the ambiguities surrounding this topic. One potential solution to these challenges lies in the exploration of Casimir energy stemming from fluctuations in various fields within compact extra dimensions. Greene and Levin \cite{Greene.} have proposed that by carefully selecting the total Casimir energy, it may be feasible to stabilize the size of the extra dimensions and propel the accelerated expansion of the three non-compact directions, with Casimir energy effectively acting as dark energy. However, it has been highlighted in references \cite{Ref.1,Ref.2,Ref.3} that the destabilization of the extra dimension occurs when contributions from matter content are considered. The notion that the acceleration of the universe can be better explained through the Casimir effect rather than through matter contributions has garnered support. Furthermore, in a study detailed in Ref. \cite{0905.0328}, the zero-order Casimir energy for free bosonic and fermionic quantum fields interacting with the {\ae}ther was computed, offering insights into stabilizing the extra dimension through a conceptual toy model. The authors in the paper claim that the non-vanishing vacuum expectation value of the {\ae}ther field would impact the dynamical equation of the background moduli field. Additionally, the Casimir energy of massless and massive fields embedded in five-dimensional space serves as dark energy and drives the expansion of the non-compact space, as expected. In this study, by expanding their issue to a Lagrangian wherein a self-interacting scalar field coupling with {\ae}ther within the fifth compact direction of space-time, we computed the radiative and thermal correction to the Casimir energy for this bosonic quantum field within an extra dimension. We also derive the thermal correction to the Casimir energy for a Dirac fermionic field residing in the fifth compact dimension of spacetime. Ultimately, with the inclusion of these new correction terms for the Casimir energy—incorporating both thermal and radiative components—we discuss the behavior of the total Casimir energy density in the compact dimension through a series of graphical representations. Concerning the radiative correction, a key aspect of our work is the specific renormalization scheme selected for our computations. Our approach to renormalization was based on a method in which counterterms are systematically acquired, considering the influences of imposed boundary conditions or background effects. In contrast to many standard renormalization programs that focus solely on renormalizing the bare parameters within the Lagrangian without consideration for field boundary conditions\cite{free.counterterm.1,free.counterterm.2}, our strategy prioritizes the consistency of counterterms with the pertinent boundary conditions or background features. We contend that in conventional renormalization programs, the alignment of counterterms with boundary conditions or background effects is often overlooked or inadequately influenced. Our perspective posits that the role of the counterterm is essential for the renormalization of the Lagrangian’s bare parameters and should inherently align with the field’s boundary conditions or background characteristics \cite{pos.dep.counterterm.}. A failure to make this selection judiciously may result in unresolved divergences in physical quantities, such as the Casimir energy. Therefore, utilizing a counterterm that accounts for the associated boundary conditions provides a more reliable assurance of the accuracy of the results. Building upon this concept, our paper delves into the systematic extraction of vacuum energy within extra dimensions for a self-interacting scalar field coupled with the {\ae}ther.  Subsequently, by applying established regularization techniques, we proceed to calculate the radiative correction to the Casimir energy for the scalar field subject to periodic boundary conditions along the circular extra dimension. Our regularization process involves a slight modification of the box subtraction scheme to navigate the complexities of infinities. Introducing a secondary system akin to the original setup in our regularization technique furnishes additional parameters that serve as regulators, thereby enhancing the removal of divergent elements with greater clarity. Ultimately, the investigation examines the total Casimir energy density in a bulk system by taking into account the radiative correction to the Casimir energy, particularly within the context of a toy model. To gauge the impact of correction terms (radiative and thermal corrections) on the total Casimir energy, we calculated the ratio of these correction values relative to the total Casimir energy. The findings were then analyzed and presented graphically to illustrate the significance of these corrections. The structure of our paper is as follows: In the next section, we introduce the model and elucidate the process of determining counterterms and vacuum energy. Subsequently, in Section \ref{Sec.Casimir.En.}, after reviewing the calculation of the leading-order Casimir energy, we provide a detailed examination of the computation process for both radiative and thermal corrections to the Casimir energy. Finally, the sign and magnitude of the total Casimir energy density are discussed through graphical representations. The concluding section provides a summary of the results and examines their broader physical implications.

\section{The model}\label{Sec:model}
The Lagrangian for a self-interacting scalar field, including the minimal coupling term with the {\ae}ther, is expressed as:
\begin{eqnarray}\label{Lagrangian.}
  \mathcal{L}=\frac{1}{2}\Big[\partial_a\phi\partial^a\phi-m_0^2\phi^2+\mu_{\phi}^{-2}u^a u^b\partial_a\phi\partial_b\phi\Big]-\frac{\lambda_0}{4!}\phi^4,
\end{eqnarray}
where $\mu_\phi$ is the coupling parameters\cite{0905.0328}. In this Lagrangian, the {\ae}ther is defined by a space-like vector field with a non-vanishing component along the extra fifth dimension.  Specifically, the vector is determined as $u=(0,0,0,0,v)$. Additionally, the parameters $m_0$ and $\lambda_0$ represent the bare mass and coupling constant, respectively. The following metric describes the 5-dimensional spacetime obtained by augmenting a circle $S^1$ to the 4-dimensional FRW-type spacetime,
\begin{eqnarray}\label{metric.Def}
       ds^2=dt^2-a(t)^2dx^idx^j\delta_{ij}-b(t)^2 d\theta^2,
\end{eqnarray}
wherein $i$ and $j$ range from $1$ to $3$, inclusive, and $a(t)$ represents the scale factor of the non-compact three-dimensional space. Additionally, the parameter $b(t)$ signifies the radius of the compact fifth direction, while the coordinate $\theta$ ranges from 0 to $2\pi$, residing on the circle $S^1$. For the scenario of a non-interacting scalar field, characterized by $\lambda_0=0$, the equation of motion is expressed as follows:
\begin{eqnarray}\label{equation.O.motion}
     \Big[\partial_a\partial^a+\mu_{\phi}^{-2}(u.\partial)^2+m^2\Big]\phi(x)=0.
\end{eqnarray}
After substituting the aforementioned {\ae}ther field as $u=(0,0,0,0,v)$ and solving the equation of motion provided in Eq. (\ref{equation.O.motion}), the dispersion relation is derived as follows:
\begin{eqnarray}\label{dispersion.relation}
     \omega_n^2=k_1^2+k_2^2+k_3^2+\frac{n^2}{b^2}(1+\alpha^2)+m^2,
\end{eqnarray}
where $\alpha=\frac{v}{\mu_\phi}$ and $n\in\mathbb{Z}$. Furthermore, the scalar field adheres to the following periodic boundary condition on the compact dimension:
\begin{eqnarray}\label{P.B.C.}
     \phi(t,x_1,x_2,x_3,\theta)= \phi(t,x_1,x_2,x_3,\theta+2\pi).
\end{eqnarray}
To derive the vacuum energy in the first order of the coupling constant $\lambda$, we require the expression for the Green's function. Thus, following the standard procedure for obtaining the Green's function expression and applying the Wick rotation, we arrive at:
\begin{eqnarray}\label{Greens.function.}
      G(x,x')=\frac{1}{\pi b}\int\frac{d\omega}{2\pi}\int\frac{d^3\mathbf{k}}{(2\pi)^3}\sum_{n=0}^{\infty}\frac{e^{-\omega(t-t')}e^{i\mathbf{k}.(\mathbf{x}-\mathbf{x}')}\cos(n\theta)\cos(n\theta')}
      {\omega^2+\omega_n^2},
\end{eqnarray}
where $\mathbf{x}=(x_1,x_2,x_3)$ and $\mathbf{k}=(k_1,k_2,k_3)$. The bare parameters $m_0$ and $\lambda_0$ in the Lagrangian, as specified in Eq. (\ref{Lagrangian.}), necessitate renormalization. To accomplish this, we introduce a rescaling of the scalar field with the parameter $Z=\delta_z+1$, commonly referred to as the field strength parameter. This rescaling operation transforms the Lagrangian of Eq. (\ref{Lagrangian.}) into:
\begin{eqnarray}\label{Lagrangian.A.Renorm.}
      \mathcal{L}&=&\frac{1}{2}\partial_a\phi_r\partial^a\phi_r-\frac{1}{2}m^2\phi_r^2-\frac{\lambda}{4!}\phi_r^4+\frac{1}{2}\mu_{\phi}^{-2}u^a u^b\partial_a\phi_r\partial_b\phi_r\nonumber\\&&+\frac{1}{2}\delta_z\partial_a\phi_r\partial^a\phi_r-\frac{1}{2}\delta_m\phi_r^2
      -\frac{\delta_\lambda}{4!}\phi_r^4+\frac{1}{2}\delta_z\mu_{\phi}^{-2}u^a u^b\partial_a\phi_r\partial_b\phi_r.
\end{eqnarray}
Here, $\delta_m=Zm_0^2-m^2$ and $\delta_\lambda=Z^2\lambda_0-\lambda$ represent the mass and coupling constant counterterms, respectively. The Feynman rules associated with these counterterms are:
\begin{eqnarray}\label{Fynmann.Rule.Counterterms}
   \raisebox{0mm}{\includegraphics[width=1cm]{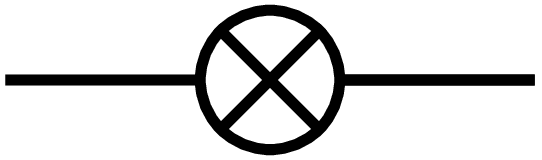}}&=&i\big[\mathcal{P}^2\delta_Z-\delta_m\big],\nonumber\\
   \raisebox{-2mm}{\includegraphics[width=0.7cm]{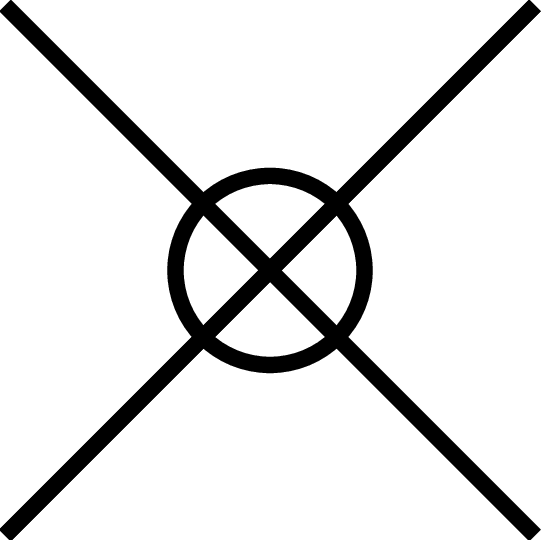}}&=&-i\delta_\lambda.
\end{eqnarray}
where $\mathcal{P}^2=p^2+\mu_{\phi}^{-2}(u.p)^2$. To calculate the counterterms $\delta_m$ and $\delta_z$ up to the first order of the coupling constant $\lambda$, we begin with the two-point function given by the following expression:
\begin{eqnarray}\label{Two.point.function}
     \raisebox{-3mm}{\includegraphics[width=1.2cm]{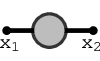}}=\raisebox{-1.5mm}{\includegraphics[width=1.2cm]{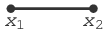}}
   +\raisebox{-2.5mm}{\includegraphics[width=1cm]{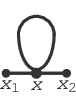}}+\raisebox{-2.5mm}{\includegraphics[width=1.2cm]{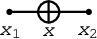}}\hspace{0cm}.
\end{eqnarray}
The standard renormalization condition necessary for determining the counterterms is:
\begin{eqnarray}\label{Renorm.Condition.}
    \raisebox{-2mm}{\includegraphics[width=1cm]{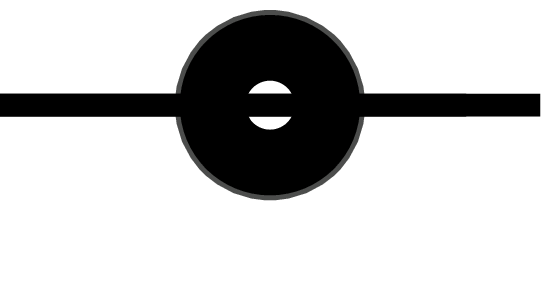}}&=&\frac{i}{\mathcal{P}^2-m^2}+\mbox{\tiny(the terms regular at $\mathcal{P}^2=m^2$)},\nonumber\\
   \raisebox{-2mm}{\includegraphics[width=0.7cm]{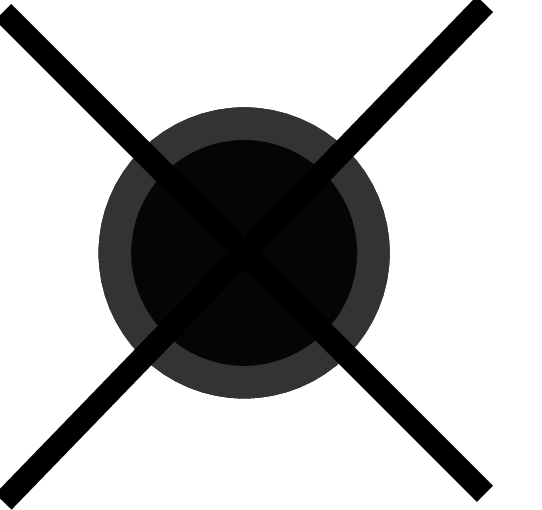}}&=&-i\lambda\hspace{2cm} \mbox{(at $s=4m^2$, $t=u=0$)}.
\end{eqnarray}
The parameters $s$, $t$, and $u$ denote the type of channel. As is well-known, the channel can be inferred from the form of the Feynman diagram, with each channel leading to characteristic angular dependencies of the cross-section. By utilizing the two-point function provided in Eq. (\ref{Two.point.function}) and the renormalization condition expressed in Eq. (\ref{Renorm.Condition.}), we find that the counterterms $\delta_z$ and $\delta_\lambda$ are zero up to the first order of the coupling constant $\lambda$, while the mass counterterm $\delta_m$ has a non-zero value, determined as follows:
\begin{eqnarray}\label{mass.counterterm.}
     \delta_m(x)=\frac{-i}{2}\raisebox{-2mm}{\includegraphics[width=1cm]{15.eps}}=\frac{-\lambda}{2}G(x,x),
\end{eqnarray}
where the function $G(x,x')$ is the Green's function expression. The positional dependence of the counterterm, as evident from its relation, provides a mechanism to integrate the effects of spatial distortion into the counterterm expression. This dependency facilitates the incorporation of boundary conditions or non-trivial backgrounds. It is note of that, the use of position-dependent counterterms to renormalize the bare parameters of the Lagrangian is not a new concept and it has been extensively explored in prior research\cite{koli.1,koli.2,koli.3}. The difference lies in how these counterterms are applied in the renormalization process. For instance, authors often utilized free counterterms (\emph{e. g.} the type of counterterm which is usually for Minkowski space-time) in the space between boundaries, resorting to position-dependent counterterms near the boundaries\cite{freevapositioncounterterm,fosco.}. However, in this paper, inspired by the pioneering work of S.S. Gousheh \emph{et al.}, we adopt a systematically self-consistent approach to derive position-dependent counterterms\cite{2D.Man}. Here, the Green's function plays a crucial role in capturing all effects arising from the boundary conditions. Consequently, it can be inferred that the effects of boundary conditions are encapsulated in the counterterm via the Green's function. Consequently, the expression for vacuum energy takes the following form:
\begin{eqnarray}\label{Vacuum.En.EXP1.}
     E^{(1)}_{\mbox{\tiny vac.}}&=&i\int_{V} \sqrt{-g}d^4x\bigg(\frac{1}{8} \raisebox{-7mm}{\includegraphics[width=0.5cm]{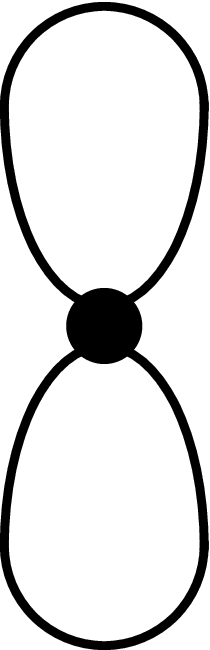}}
       +\frac{1}{2}\raisebox{-1mm}{\includegraphics[width=0.5cm]{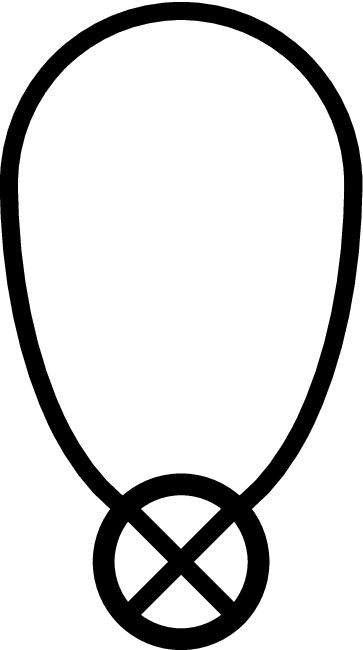}}+...\bigg)
       \\&=&i \int_{V}\sqrt{-g}d^4x\bigg(\frac{-i\lambda}{8}G^2(x,x)-\frac{-i}{2}\delta_m(x)G(x,x)\bigg),\nonumber
\end{eqnarray}
Here, $g$ denotes the determinant of the metric tensor $g_{\mu\nu}$, which is derived from the line element expression provided in Eq. (\ref{metric.Def}). Furthermore, the parameter $V$ represents a 4-dimensional volume comprising a box with a specific size (\emph{e.g.}, size $L$) and a circle with radius $b$ in the compact dimension. The superscript $(1)$ on the vacuum energy denotes the first order of the coupling constant $\lambda$. Substituting Eq. (\ref{mass.counterterm.}) into Eq. (\ref{Vacuum.En.EXP1.}) results in:
\begin{eqnarray}\label{Vacuum.En.EXP2.}
  E^{(1)}_{\mbox{\tiny vac.}}=\frac{-\lambda}{8}\int_{V}\sqrt{-g}d^4x G^2(x,x).
\end{eqnarray}
\section{Casimir Energy}\label{Sec.Casimir.En.}
\begin{figure}[th]\centering\includegraphics[width=9cm]{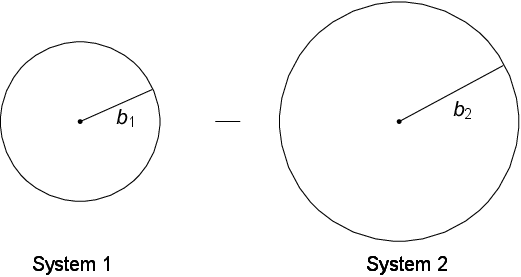}
\caption{\label{PLOT.Sphere.BSS}
The left figure illustrates a circle with a radius \(b_1\), referred to as ``System 1''. In contrast, the right figure depicts another circle with a larger radius \(b_2 > b_1\), called ``System 2.'' To compute the Casimir energy using the BSS method, you need to subtract the vacuum energies of these two configurations. This is done following the process outlined in Eq. (\ref{BSS.Def}).}
\end{figure}

In this section, we begin with a detailed computational analysis of the leading-order and next-to-leading-order Casimir energy for a massive scalar field coupled with {\ae}ther in a circular compact dimension. Following this, we calculate the thermal correction to the Casimir energy for massive and massless bosons and fermions. To achieve this, we introduce two analogous systems with distinct circle radii, denoted as $b_1$ and $b_2$, respectively, within the compact dimension. The Casimir energy is computed by taking the difference in vacuum energies between these two systems. Subsequently, to obtain the Casimir energy expression for the original system (with radius $b_1$), we allow the radius associated with the second system (with radius $b_2$) to tend towards infinity. We assert that as the size of the second system approaches infinity, its vacuum energy converges to that of a system without any boundary conditions. This computational method is recognized in the literature as a regularization technique, termed the ``Box Subtraction Scheme'', which we refer to as BSS for brevity. The BSS facilitates the inclusion of additional parameters in the vacuum energy subtraction process, serving as regulators to enhance the clarity of divergence elimination. The Casimir energy definition based on the BSS is typically expressed by the following relation:
\begin{eqnarray}\label{BSS.Def}
      E_{\mbox{\tiny Cas.}}=\lim_{b_2\to\infty}\Big[\Delta E^{(0)}_{\mbox{\tiny Vac.}}+\Delta E^{(1)}_{\mbox{\tiny Vac.}}+\Delta E^{(T)}_{\mbox{\tiny Vac.}}\Big].
\end{eqnarray}
Here, $\Delta E^{(0)}_{\mbox{\tiny Vac.}}$ represents the subtraction of zero-point energies between the two systems depicted in Fig. (\ref{PLOT.Sphere.BSS}), while $\Delta E^{(1)}_{\mbox{\tiny Vac.}}$ denotes the subtraction of vacuum energies for the next-to-leading-order case (refer to Eq. (\ref{Vacuum.En.EXP2.}) for the vacuum energy expression). Moreover, the term $\Delta E^{(T)}_{\mbox{\tiny Vac.}}$ represents the subtraction of thermal corrections from the vacuum energy between the two systems illustrated in Fig.  (\ref{PLOT.Sphere.BSS}). In the following subsections, we compute and present detailed calculations for the Casimir energy corresponding to each correction term.
\subsubsection{Leading-Order Casimir Energy}\label{Subsec:: ZeroOrder}
Initially, we present a comprehensive analysis concerning the leading-order Casimir energy for a massive scalar field coupled with {\ae}ther in a circular compact dimension. The result for the leading-order Casimir energy of such a system has been previously reported in \cite{0905.0328}. However, by employing the BSS method to compute this order of Casimir energy, we offer an alternative perspective on this matter, hereby enhancing confidence in the efficacy of this regularization technique. So, by starting with zero point energy density expression, we have
\begin{eqnarray}\label{Zero-Order Cas. EXPR1}
     \Delta \mathcal{E}^{(0)}_{\mbox{\tiny Vac.}}=\frac{1}{2(2\pi b_1 L^3)}\int\frac{L^3d^3\mathbf{k}}{(2\pi)^3}\sum_{n\in\mathbb{Z}}\Big[\mathbf{k}^2+\frac{n^2}{b_1^2}(1+\alpha^2)+m^2\Big]^{\frac{1}{2}}-\{b_1\to b_2\},
\end{eqnarray}
where $\mathbf{k}=(k_1,k_2,k_3)$. To regularize the summation over $n$, we convert it into integral form. This conversion is typically achieved using the Abel-Plana Summation Formula\,(APSF), defined as follows:
\begin{eqnarray}\label{APSF.definition}
     \sum_{n=0}^{\infty}\mathcal{F}(n)=\frac{1}{2}\mathcal{F}(0)+\int_{0}^{\infty}\mathcal{F}(x)dx+i\int\frac{\mathcal{F}(it)-\mathcal{F}(-it)}{e^{2\pi t}-1}dt.
\end{eqnarray}
In the APSF, the first term on the right-hand side of Eq. (\ref{APSF.definition}) is typically known as the \emph{zero-term}, the second term is referred to as the \emph{integral-term}, and the final term is commonly called the \emph{branch-cut} term\cite{APSF.Ref.1,APSF.Ref.2,APSF.Ref.3}. Combining Eqs. (\ref{Zero-Order Cas. EXPR1}) and (\ref{APSF.definition}) yields the subsequent expression,
\begin{eqnarray}\label{Zero-Order Cas. EXPR2}
      \Delta \mathcal{E}^{(0)}_{\mbox{\tiny Vac.}}=\frac{1}{(2\pi)^4 b_1}\left\{\frac{b_1 \Omega_4}{2\sqrt{1+\alpha^2}}
      \int_{0}^{\infty} \eta^3\sqrt{\eta^2+m^2}d\eta+\mathcal{B}(b_1)\right\}-\{b_1\to b_2\},
\end{eqnarray}
where the change of variables $\eta=(k_1,k_2,k_3,X)$ and $X=x\sqrt{1+\alpha^2}/b_1$ has been applied. Additionally, $\Omega_d=\frac{2\pi^{d/2}}{\Gamma(d/2)}$ represents the spatial angle. The first term within the brackets of Eq. (\ref{Zero-Order Cas. EXPR2}), which comes from the integral term of the APSF, does not depend on the radii \(b_1\) and \(b_2\). As a result, its contribution is analytically canceled out by the corresponding term in the vacuum energy of the second system. Consequently, only the branch-cut term remains from Eq. (\ref{Zero-Order Cas. EXPR2}), leading us to derive:
\begin{eqnarray}\label{Zero-Order Cas. EXPR3}
      \Delta \mathcal{E}^{(0)}_{\mbox{\tiny Vac.}}=\frac{1}{(2\pi)^4 b_1}\left\{\frac{-8\pi b_1}{\sqrt{1+\alpha^2}}\int_{m}^{\infty}d\tau\int_{0}^{\sqrt{\tau^2-m^2}}k^2dk\frac{\sqrt{\tau^2-k^2-m^2}}{e^{2\pi b_1\tau/\sqrt{1+\alpha^2}}-1}\right\}-\{b_1\to b_2\},
\end{eqnarray}
where $\tau=t\sqrt{1+\alpha^2}/b_1$. Obtaining a closed-form expression directly from the result of the above integration is arduous. Therefore, we simplify the computation of integrals over $k$ and $\tau$ by employing the relation $\sum_{j=1}^{\infty} e^{-Aj} = \frac{1}{e^A - 1}$ for the denominator of the above equation. Ultimately, by implementing the limit $b_2 \to \infty$ to derive the Casimir energy, the expression for the leading-order Casimir energy is obtained as:
\begin{eqnarray}\label{Zero.Cas.EXP.MASSIVE}
       \mathcal{E}^{(0)}_{\mbox{\tiny Cas.}}=\frac{-\left(\alpha ^2+1\right)}{(2\pi)^7}\sum_{j=1}^{\infty}\frac{ e^{-\frac{2 \pi  b_1 j m}{\sqrt{\alpha ^2+1}}} \left(3 \left(\alpha ^2+1\right)+2 \pi  b_1 j m \left(3 \sqrt{\alpha ^2+1}+2 \pi  b_1 j m\right)\right)}{ b_1^5 j^5}.
\end{eqnarray}
Returning to Eq. (\ref{Zero-Order Cas. EXPR1}) and setting the mass of the field as $m = 0$, subsequent computations of the Casimir energy, following the same procedures as those employed for the massive case, yield the Casimir energy density of the massless scalar field as the following relation:
\begin{eqnarray}\label{Zero.Cas.EXP.MASSLESS}
        \mathcal{E}^{(0)}_{\mbox{\tiny Cas.}}=\frac{-3\left(\alpha^2+1\right)^{2} \zeta(5)}{(2\pi)^7 b_1^5}.
\end{eqnarray}
In Fig. (\ref{PLOT.RCMassCheck}), a series of plots depicting the Casimir energy for the massive case were generated. The observed trend, where the behavior approaches that of the massless case for mass values \( m \leq 0.1 \), underscores the physical consistency between the results derived in Eqs. (\ref{Zero.Cas.EXP.MASSIVE}) and (\ref{Zero.Cas.EXP.MASSLESS}). The results obtained via the BSS, given in Eqs. (\ref{Zero.Cas.EXP.MASSIVE}) and (\ref{Zero.Cas.EXP.MASSLESS}), are also entirely consistent with those reported in previous works \cite{0905.0328}.
\subsubsection{Radiative Correction}\label{Subsec::RC}
\begin{figure}[th]\centering\includegraphics[width=8.5cm]{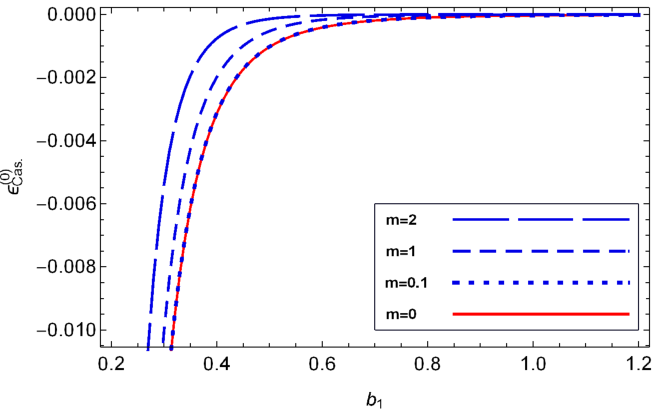}\includegraphics[width=8cm]{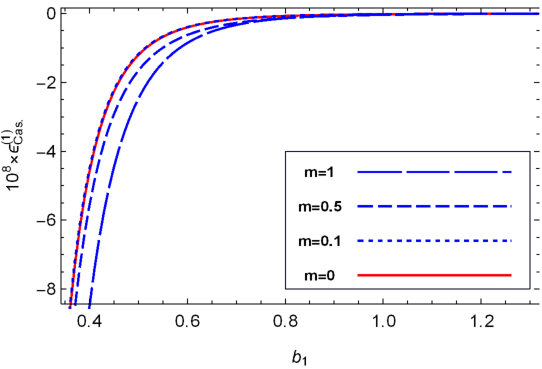}
\caption{\label{PLOT.RCMassCheck}
The left figure shows the leading-order Casimir energy for a scalar field, both massive and massless, coupled with {\ae}ther in a circular compact dimension. The energy is plotted against the radius \(b_1\) for mass values \(m = \{2, 1, 0.1, 0\}\). As the mass parameter decreases, the graphs for the massive scalar field converge with those for the massless field, demonstrating the consistency between the results derived from Eqs. (\ref{Zero.Cas.EXP.MASSIVE}) and (\ref{Zero.Cas.EXP.MASSLESS}). In the right figure, the radiative correction to the Casimir energy density for a self-interacting scalar field—whether massive or massless—coupled with an {\ae}ther in a circular compact dimension, is plotted as a function of radius \(b_1\). The plots correspond to various field masses \(m = \{1, 0.5, 0.1, 0\}\). The trend observed in these plots indicates that as the mass decreases, the radiative correction from Eq. (\ref{RC.CAS.Final.Massive}) converges toward that of the massless scalar field. The data suggests that when the mass is below \(0.1\), the difference in radiative correction between the massive and massless scalar fields becomes negligible. The coupling constant parameters used for these plots are $\lambda = 0.1$ and $\alpha = 1$. All plots are in units where $\hbar c=1$.}
\end{figure}

To derive the next-to-the-leading-order Casimir energy, we begin with the second term within the brackets of Eq. (\ref{BSS.Def}). By combining this term with Eqs. (\ref{Greens.function.}) and (\ref{Vacuum.En.EXP2.}), we obtain:
\begin{eqnarray}\label{NLO.EXP.1}
      \Delta E^{(1)}_{\mbox{\tiny Vac.}}=\frac{-\lambda L^3\pi^2}{8(\pi b_1)^2(2\pi)^8}\int_{0}^{2\pi}\bigg[\sum_{n=0}^{\infty}f_n(b_1;m)\cos^2(n\theta)\bigg]
      \bigg[\sum_{n'=0}^{\infty}f_{n'}(b_1;m)\cos^2(n'\theta)\bigg]b_1d\theta
      -\{b_1\to b_2\},
\end{eqnarray}
where the function $f_n(b_1;m)$ is
\begin{eqnarray}\label{F.Definition}
      f_n(b_1;m)=\int\frac{d^3\mathbf{k}}{\left(\mathbf{k}^2+\frac{n^2}{b_1^2}(1+\alpha^2)+m^2\right)^{1/2}}.
\end{eqnarray}
Integration over $\theta$ converts Eq. (\ref{NLO.EXP.1}) to
\begin{eqnarray}\label{NLO.EXP.2}
    \Delta E^{(1)}_{\mbox{\tiny Vac.}}=\frac{-\lambda L^3 \pi}{8(2\pi)^8}\frac{1}{b_1}\bigg[\frac{1}{2}\sum_{n,n'=1}^{\infty}f_n(b_1;m)f_{n'}(b_1;m)
    +\frac{1}{4}\sum_{n=1}^{\infty}f_n(b_1;m)^2+2f_0(b_1;m)\sum_{n=0}^{\infty}f_n(b_1;m)\bigg]
      -\{b_1\to b_2\}.
\end{eqnarray}
All summation forms presented in Eq. (\ref{NLO.EXP.2}) lead to divergent expressions. To regulate them, we utilize the APSF as defined in Eq. (\ref{APSF.definition}), converting the summation forms into integral forms. Consequently, we obtain:
\begin{eqnarray}\label{NLO.EXP.3}
     \Delta E^{(1)}_{\mbox{\tiny Vac.}}&=&\frac{-\lambda L^3 \pi}{8(2\pi)^8}\frac{1}{b_1}\bigg[
         \frac{1}{8}f_0(b_1;m)^2+\underbrace{\frac{1}{2}\left(\int_{0}^{\infty}f_x(b_1;m)dx\right)^2}_{b_1\mathcal{T}_1(b_1,\infty)}+\frac{1}{2}\mathcal{B}_1(b_1;m)^2\nonumber\\
         &&-\frac{1}{2}\underbrace{ f_0(b_1;m)\int_{0}^{\infty}f_x(b_1;m)dx}_{b_1\mathcal{T}_2(b_1,\infty)}-\frac{1}{2}\underbrace{f_0(b_1;m)\mathcal{B}_1(b_1;m)}_{b_1\mathcal{S}_1(b_1,\infty)}
         +\underbrace{\mathcal{B}_1(b_1;m)\int_{0}^{\infty}f_x(b_1;m)dx}_{b_1\mathcal{S}_2(b_1,\infty)}\nonumber\\
         &&+\frac{-1}{8}f_0(b_1;m)^2+\underbrace{\frac{1}{4}\int_{0}^{\infty}f_x(b_1;m)^2dx}_{b_1\mathcal{T}_3(b_1,\infty)}+\mathcal{B}_2(b_1;m)\nonumber\\
         &&+\underbrace{f_0(b_1;m)^2}_{b_1\mathcal{T}_4(b_1,\infty)}+2f_0(b_1;m)\int_{0}^{\infty}f_x(b_1;m)dx+2f_0(b_1;m)\mathcal{B}_3(b_1;m)\bigg]-\{b_1\to b_2\}.
\end{eqnarray}
The first term and the seventh term within the brackets of the above equation cancel each other automatically. Despite this, there are still some expressions within the bracket that remain divergent. To regulate these terms, in addition to the BSS, we employ the cutoff regularization technique. The methodology for regularizing divergent expressions and removing these infinite terms is elaborated in detail in Appendix \ref{Appendix.Finding Finite Parts}. Following the cancellations, the final result is:
\begin{eqnarray}\label{Eq.After.Appen.}
  \Delta E^{(1)}_{\mbox{\tiny Vac.}}&=&\frac{-\lambda L^3 \pi}{8(2\pi)^8}\frac{1}{b_1}\bigg[
         \frac{1}{2}\mathcal{B}_1(b_1;m)^2
         +\frac{3m^2\pi}{2}\mathcal{B}_1(b_1;m)
         +\frac{2m^3\pi^2b_1}{3\sqrt{1+\alpha^2}}\mathcal{B}_1(b_1;m)+\mathcal{B}_2(b_1;m)\bigg]-\{b_1\to b_2\}.
\end{eqnarray}
The expressions $\mathcal{B}_1(b_1;m)$ and $\mathcal{B}_2(b_1;m)$ denote the branch-cut terms of APSF, each possessing finite values. In computing the value of $\mathcal{B}_1(b_1;m)$, we begin with,
\begin{eqnarray}\label{Branchcut.1}
      \mathcal{B}_1(b_1;m)&=&i\int dt\frac{f_{it}(b_1;m)-f_{-it}(b_1;m)}{e^{2\pi t}-1}\nonumber\\&=&\frac{8\pi b_1}{\sqrt{1+\alpha^2}}\int_{m}^{\infty}dT\int_{0}^{\sqrt{T^2-m^2}}
      k^2dk\frac{\left[T^2-k^2-m^2\right]^{-1/2}}{e^{2\pi T b_1/\sqrt{1+\alpha^2}}-1},
\end{eqnarray}
wherein $T=\frac{t}{b_1}\sqrt{1+\alpha^2}$ and changing the order of integration has been applied. Subsequently, utilizing the identity $\sum_{j=1}^{\infty} e^{-Aj} = \frac{1}{e^A - 1}$, integration computations yield the following expression for the branch-cut term $\mathcal{B}_1(b_1;m)$:
\begin{eqnarray}\label{Branchcut.2}
       \mathcal{B}_1(b_1;m)=\frac{\sqrt{1+\alpha^2}}{2\pi b^2_1}
       \sum_{j=1}^{\infty}\frac{e^{-2\pi m jb_1/\sqrt{1+\alpha^2}}\left[2\pi m j b_1+\sqrt{1+\alpha^2}\right]}{j^3}.
\end{eqnarray}
A similar computation for the branch-cut term \( \mathcal{B}_2(b_1;m) \) results in a value of zero. Following the definition of the Casimir energy in BSS, as given in Eq. (\ref{BSS.Def}), the final step involves computing the limit \( b_2 \to \infty \).  Upon evaluating this limit for Eq. (\ref{Eq.After.Appen.}), all contributions, including the finite contributions associated with the second system (the system with radius $b_2$), are eliminated. Consequently, the ultimate expression for the radiative correction to the Casimir energy density for a massive scalar field coupled with {\ae}ther residing in the fifth compact dimension is obtained as follows:
\begin{eqnarray}\label{RC.CAS.Final.Massive}
       \mathcal{E}^{(1)}_{\mbox{\tiny Cas.}}&=&\frac{-\lambda \pi}{8(2\pi)^9b_1^2}\bigg[
         \frac{1}{2}\mathcal{B}_1(b_1;m)
         +\frac{3m^2\pi}{2}
         +\frac{2m^3\pi^2b_1}{3\sqrt{1+\alpha^2}}\bigg]\mathcal{B}_1(b_1;m).
\end{eqnarray}
As the mass parameter approaches zero in Eq. (\ref{RC.CAS.Final.Massive}), it leads to the radiative correction to the Casimir energy density for the massless scalar field coupled with {\ae}ther. The resulting expression following this limit is:
\begin{eqnarray}\label{RC.CAS.Final.Massless}
       \mathcal{E}^{(1)}_{\mbox{\tiny Cas.}}=\frac{-\lambda \left(1+\alpha ^2\right)^2 \zeta (3)^2}{32768 \pi ^{10} b_1^6}.
\end{eqnarray}
In Fig. (\ref{PLOT.RCMassCheck}), the radiative correction to the Casimir energy density for a massive scalar field coupled with {\ae}ther is depicted as \( b_1 \), representing the radius of the circle defined in the fifth compact dimension. This figure illustrates that as the mass value decreases, the result for the radiative correction approaches that of the massless case.

\subsubsection{Thermal Correction}\label{Subsec:: ThermalCorrection}
\begin{figure}[th]\centering\includegraphics[width=8cm]{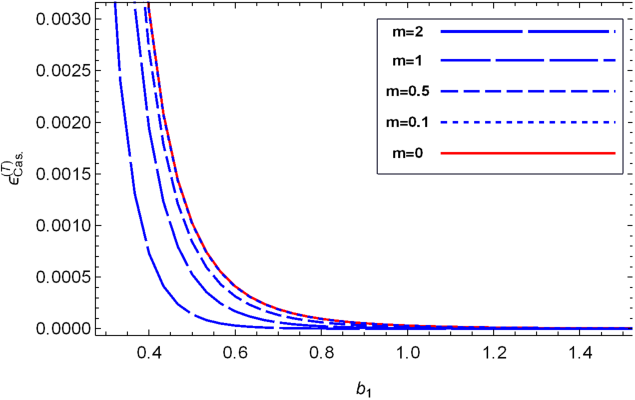}
\includegraphics[width=8cm]{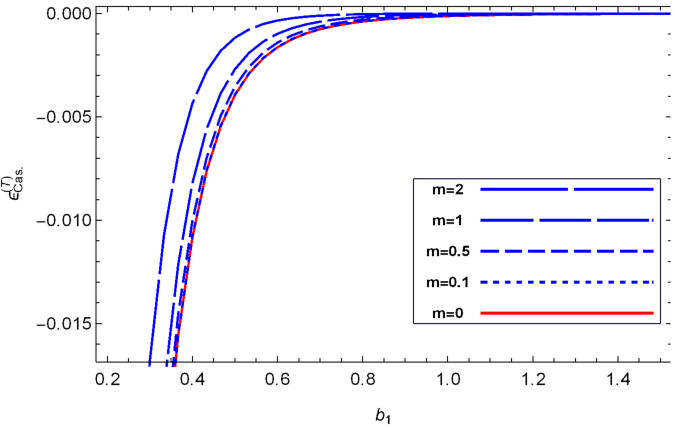}
\caption{\label{PLOT.ThermalMassCheck}
In the left (right) figure, the thermal correction to the Casimir energy density for a massive/massless scalar field (Dirac fermion field) coupled with {\ae}ther on a circular compact dimension is plotted as a function of radius \(b_1\). These figures show results for a sequence of field masses \(m = \{1, 0.5, 0.1, 0\}\). The trend in the plots indicates that as the mass decreases, the thermal correction term rapidly converges to the value for the massless field. This convergence is observed in both scalar fields and Dirac fermion fields. The coupling constant parameters used for these plots are $\lambda= 0.1$, $\alpha=1$, and $\alpha_f=1$. All plots are in units where $\hbar c=1$.}
\end{figure}

The thermal correction for vacuum energy is typically represented by the following expression:
\begin{eqnarray}\label{ThermalEq.1}
      E^{(T)}=\sum_{n} \frac{\epsilon\omega_n}{e^{\beta\omega_n}-\epsilon},
\end{eqnarray}
wherein the parameter \(\epsilon\) takes the value \(1\) for bosons and \(-1\) for fermions. To calculate the thermal correction to the Casimir energy density for a massive scalar field existing within the fifth compact dimension of spacetime, we begin with the third term in the bracket of Eq. (\ref{BSS.Def}). By using the dispersion relation outlined in Eq. (\ref{dispersion.relation}), the following expression can be derived:
\begin{eqnarray}\label{ThermalEq.2}
      \Delta \mathcal{E}_{\mbox {\tiny vac.}}^{(T)}=\frac{1}{(2\pi)^4 b_1}\int d^3\mathbf{k}\sum_{j=1}^{\infty}\sum_{n\in \mathbb{Z}} \left[\mathbf{k}^2+\frac{n^2}{b_1^2}(1+\alpha^2)+m^2\right]^{1/2}
       e^{-j\beta\left[\mathbf{k}^2+\frac{n^2}{b_1^2}(1+\alpha^2)+m^2\right]^{1/2}}
      -\{b_1\to b_2\},
\end{eqnarray}
where we used the identity \(\frac{1}{e^A - 1} = \sum_{j = 1}^{\infty} e^{-jA}\). Here, \(\mathbf{k} = (k_1, k_2, k_3)\), and \(\alpha = v/\mu_\phi\). To regularize the summation in this expression, we applied the APSF as defined in Eq. (\ref{APSF.definition}), transforming the summation into an integral. This leads to the following result:
\begin{eqnarray}\label{Thermal.Eq.After.APSF}
      \Delta \mathcal{E}_{\mbox {\tiny vac.}}^{(T)}=\frac{\Omega_4}{(2\pi)^4\sqrt{1+\alpha^2}}\sum_{j=1}^{\infty}\int_{0}^{\infty}\xi^3 \left[\xi^2+m^2\right]^{1/2}e^{-j\beta\left[\xi^2+m^2\right]^{1/2}}d\xi+\mathcal{B}_{\mbox{\tiny T}}(b_1,m)-\{b_1\to b_2\}.
\end{eqnarray}
Here, we used the variable substitution $\xi = (\mathbf{k}, \frac{x}{b_1} \sqrt{1 + \alpha^2})$ for the first term (the integral term) of the preceding equation. As indicated in Eq. (\ref{Thermal.Eq.After.APSF}), the first term is independent of the radii \(b_1\) and \(b_2\), and hence, its contribution is analytically canceled by similar terms derived from the second system (the system with a radius of \(b_2\)). Consequently, the only remaining terms from Eq. (\ref{Thermal.Eq.After.APSF}) are the branch-cut terms \(\mathcal{B}_{\mbox{\tiny T}}(b_1, m)\) and \(\mathcal{B}_{\mbox{\tiny T}}(b_2, m)\). Therefore, the resulting expression is:
\begin{eqnarray}\label{Thermal.Only Branch}
       \Delta \mathcal{E}_{\mbox {\tiny vac.}}^{(T)}=\mathcal{B}_{\mbox{\tiny T}}(b_1,m)-\mathcal{B}_{\mbox{\tiny T}}(b_2,m)=2i\sum_{j=1}^{\infty}\int \frac{dt}{e^{2\pi t}-1}\Big[\mathcal{U}_j(it;b_1,m)-\mathcal{U}_j(-it;b_1,m)\Big]-\{b_1\to b_2\}
\end{eqnarray}
where
\begin{eqnarray}\label{U}
      \mathcal{U}_j(n;b_1,m)=\frac{1}{(2\pi)^4 b_1}\int d^3\mathbf{k}
      \left[\mathbf{k}^2+\frac{n^2}{b_1^2}(1+\alpha^2)+m^2\right]^{1/2}e^{-j\beta \left[\mathbf{k}^2+\frac{n^2}{b_1^2}(1+\alpha^2)+m^2\right]^{1/2}}.
\end{eqnarray}
By changing the variable to $T = t \sqrt{1 + \alpha^2} / b_1$, setting \(p^2 = \mathbf{k}^2 + m^2\), and then altering the variable to \(x^2 = T^2 / p^2 - 1\), the expression of Eq. (\ref{Thermal.Only Branch}) is transformed into the following form:
\begin{eqnarray}\label{Thermal.Eq. after.U}
       \Delta \mathcal{E}_{\mbox {\tiny vac.}}^{(T)}=\frac{-8}{(2\pi)^3\sqrt{1+\alpha^2}}\sum_{j=1}^{\infty}\int_{0}^{\infty}dx\int_{m}^{\infty}dp\frac{x^2p^3(p^2-m^2)^{1/2}\cos(j\beta p x)}{\sqrt{x^2+1}\left[e^{2\pi b_1p\sqrt{x^2+1}/{\sqrt{1+\alpha^2}}}-1\right]}-\{b_1\to b_2\}.
\end{eqnarray}
After changing the variable to $z = 2\pi b_1 p / \sqrt{1 + \alpha^2}$, and utilizing the identity \(1 / (e^{z \sqrt{x^2 + 1}} - 1) = \sum_{\tilde{n} = 1}^{\infty} e^{-\tilde{n} z \sqrt{x^2 + 1}}\), the integration over $x$ leads to the following result:
\begin{eqnarray}\label{Thermal.Eq. after.U.1}
       \Delta \mathcal{E}_{\mbox {\tiny vac.}}^{(T)}&=&\frac{-8}{(2\pi)^3\sqrt{1+\alpha^2}}\sum_{\tilde{n},j=1}^{\infty}\frac{\partial}{\partial B_j}B_j A_{j,\tilde{n}}^5\int_{M_{j,\tilde{n}}}^{\infty} \eta^2\left[\eta^2-M_{j,\tilde{n}}^2\right]^{1/2}K_1(\eta)d\eta-\{b_1\to b_2\},
\end{eqnarray}
where $A_{j,\tilde{n}}=\frac{\sqrt{1+\alpha^2}}{2\pi b_1\sqrt{B_j^2+\tilde{n}^2}}$, $M_{j,\tilde{n}}=\frac{m}{A_{j,\tilde{n}}}$ and $B_j=j\beta\frac{\sqrt{1+\alpha^2}}{2\pi b_1}$. Furthermore, the variable change \(\eta = z \sqrt{B_j^2 + \tilde{n}^2}\) has been applied. Next, the integral over \(\eta\) is calculated. Ultimately, by taking the limit as \(b_2 \to \infty\), as shown in Eq. (\ref{BSS.Def}), the final expression for the thermal correction of the Casimir energy for a massive scalar field coupled with {\ae}ther in the fifth dimension of spacetime is obtained as follows:
\begin{eqnarray}\label{final.EXP.Thermal.MASSIVE.BOSON}
      \mathcal{E}_{\mbox {\tiny Cas.B}}^{(T)}=\frac{-1}{2\pi^2\sqrt{1+\alpha^2}}\sum_{\tilde{n},j=1}^{\infty}\frac{\partial}{\partial B_j}\left[B_j A_{j,\tilde{n}}^5
      \big(3+3M_{j,\tilde{n}}+M_{j,\tilde{n}}^2\big)e^{-M_{j,\tilde{n}}}\right].
\end{eqnarray}
To derive the expression for the thermal correction to the Casimir energy for massless scalar field coupled with {\ae}ther, you can set the parameter mass $m$ to 0 in Eq. (\ref{final.EXP.Thermal.MASSIVE.BOSON}). This leads to the following expression:
\begin{eqnarray}\label{final.EXP.Thermal.MASSLESS.BOSON}
        \mathcal{E}_{\mbox {\tiny Cas.B}}^{(T)}\buildrel {m\to0}\over\longrightarrow\frac{-3(1+\alpha^2)^2}{64\pi^7b_1^5}\sum_{\tilde{n},j=1}^{\infty}\frac{\partial}{\partial B_j}\Big[\frac{B_j}{(B_j^2+\tilde{n}^2)^{5/2}}\Big].
\end{eqnarray}
The expressions for the thermal correction to the Casimir energy for both massive and massless scalar fields, as described in Eqs. (\ref{final.EXP.Thermal.MASSIVE.BOSON}) and (\ref{final.EXP.Thermal.MASSLESS.BOSON}), were plotted in Fig. (\ref{PLOT.ThermalMassCheck}). This figure illustrates the expected consistency between the massive and massless cases, confirming the physical basis underlying the results. To derive the thermal correction to the Casimir energy for a Dirac fermionic field coupled with {\ae}ther, we start by presenting the Lagrangian. As is commonly known\cite{Fermion.Lagrangian}, the Lagrangian for a fermionic field with a minimal coupling term with {\ae}ther is given by:
\begin{eqnarray}\label{Lagrangian.FERMION}
      \mathcal{L}_F=i\bar{\psi}\gamma^a\partial_a\psi-m_f\bar{\psi}\psi-\frac{i}{\mu_\psi^2}u^au^b\bar{\psi}\gamma_a\partial_b\psi,
\end{eqnarray}
where \(\mu_\psi\) represents the fermionic coupling constant with {\ae}ther, and \(m_f\) denotes the mass of the fermion. Similarly to the scalar field case, by applying the equations of motion corresponding to the Lagrangian provided in Eq. (\ref{Lagrangian.FERMION}) and using Eq. (\ref{metric.Def}), and then imposing a periodic boundary condition on the fifth dimension of spacetime, the dispersion relation for the fermionic case can be derived as follows:
\begin{eqnarray}\label{dispersion.relation.FERMION}
        \omega_n^2=k_1^2+k_2^2+k_3^2+\frac{n^2}{b^2}(1+\alpha_\psi^2)^2+m_f^2.
\end{eqnarray}
Here, $\alpha_\psi = v / \mu_\psi$, with $n \in \mathbb{Z}$. Using Eq. (\ref{ThermalEq.1}), with \(\epsilon = -1\), and repeating the same computation process as detailed in Subsection (\ref{Subsec:: ThermalCorrection}), the resulting expression for the thermal correction to the Casimir energy density for the massive fermionic field coupled with {\ae}ther is as follows:
\begin{eqnarray}\label{final.EXP.Thermal.MASSIVE.FERMION}
       \mathcal{E}_{\mbox {\tiny Cas.F}}^{(T)}=\frac{1}{2\pi^2(1+\alpha_\psi^2)}\sum_{\tilde{n},j=1}^{\infty}\frac{\partial}{\partial \mathcal{B}_j}\left[\mathcal{B}_j\mathcal{A}_{j,\tilde{n}}^5
      \big(3+3\mathcal{M}_{j,\tilde{n}}+\mathcal{M}_{j,\tilde{n}}^2\big)e^{-\mathcal{M}_{j,\tilde{n}}}\right],
\end{eqnarray}
where $\mathcal{A}_{j,\tilde{n}}=\frac{1+\alpha_\psi^2}{2\pi b_1\sqrt{\mathcal{B}_j^2+\tilde{n}^2}}$, $\mathcal{M}_{j,\tilde{n}}=\frac{m_f}{\mathcal{A}_{j,\tilde{n}}}$ and $\mathcal{B}_j=j\beta\frac{1+\alpha_\psi^2}{2\pi b_1}$. By setting the mass of the fermionic field to \(m_f = 0\) in Eq. (\ref{final.EXP.Thermal.MASSIVE.FERMION}), the thermal correction to the Casimir energy density for massless fermions is given by the following expression:
\begin{eqnarray}\label{final.EXP.Thermal.MASSLESS.FERMION}
      \mathcal{E}_{\mbox {\tiny Cas.F}}^{(T)}\buildrel {m\to0}\over\longrightarrow\frac{3(1+\alpha_\psi^2)^4}{64\pi^7b_1^5}\sum_{\tilde{n},j=1}^{\infty}\frac{\partial}{\partial \mathcal{B}_j}\Big[\frac{\mathcal{B}_j}{(\mathcal{B}_j^2+\tilde{n}^2)^{5/2}}\Big]
\end{eqnarray}
\begin{figure}[th]\centering\includegraphics[width=8cm]{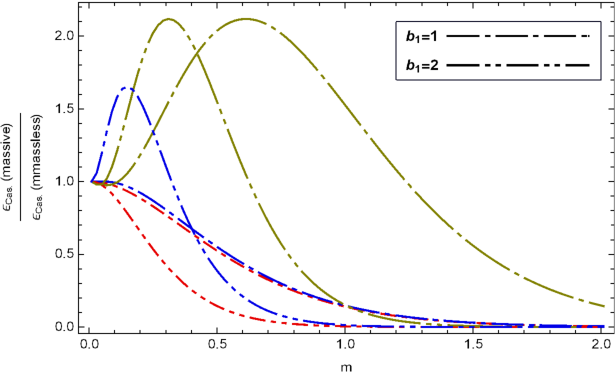}
\includegraphics[width=8cm]{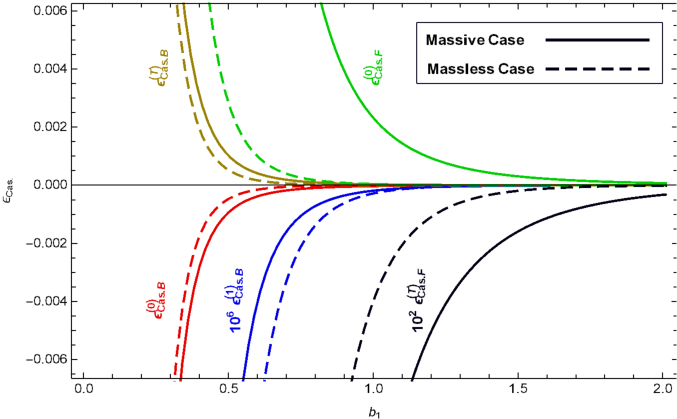}
\caption{\label{PLOT.RatioAsMass.AllPlot}
In the left figure, the ratio of the Casimir energy for a massive scalar field coupled with the {\ae}ther on a circular compact dimension, compared to the massless case, is plotted as a function of the field's mass \( m \). This figure shows the results for two distinct values of the compact dimension radius, \( b_1 = \{1, 2\} \). Three types of dashing are observed in the plots: the shortest dashing (red) represents the ratio of the zero-order Casimir energy for the massive and massless cases; the moderately dashed lines (blue) correspond to the thermal correction term; and the long dashing (brown) indicates the radiative correction term. In the right figure, the Casimir energy of both massive and massless scalar fields (Dirac fields) is plotted as a function of the radius of the compact dimension. Separate plots are provided for each order of energy, including zero-order, radiative correction, and thermal correction. This figure facilitates a direct comparison of the Casimir energy contributions from each correction term side by side. It helps visually assess the significance of each term's contribution to the total energy. The values of the mass in the right figure are set as \( m = m_f = 1 \). The coupling constant parameters used for these plots are \( \lambda = 0.1 \), \( \alpha = 1 \), and \( \alpha_f = 1 \). All plots are in units where \( \hbar c = 1 \).}
\end{figure}
\begin{table}
  \centering
  \begin{tabular}{|c||c|c|c|}
  \hline
  Particles & Degrees of freedom(Number of bosons/fermions) & Mass & Coupling constant \\ \hline\hline
  a massless bulk scalar field  & $N_b=5$ & 0 & $\alpha$\\
  a massless bulk fermion field & $N_f=8$ & 0 & $\alpha_f$\\
  a massive bulk scalar field & $\mathcal{N}_b=8$ & $m$ &$\alpha$\\
  a massive bulk fermion field & $N_f=8$ & $m_f$& $\alpha_f$\\
  \hline
\end{tabular}
  \caption{ The particle spectrum in the bulk, their degrees of freedom, their mass and the coupling constants characterizing their interaction with the {\ae}ther field. For simplicity, we assume the
  universal fermionic coupling $\alpha_f$ for both massless and massive Dirac fermion and univesal coupling constant $\alpha$ for both massless and massive boson.
}\label{table}
\end{table}
\par
\begin{figure}[th]\centering\includegraphics[width=8cm]{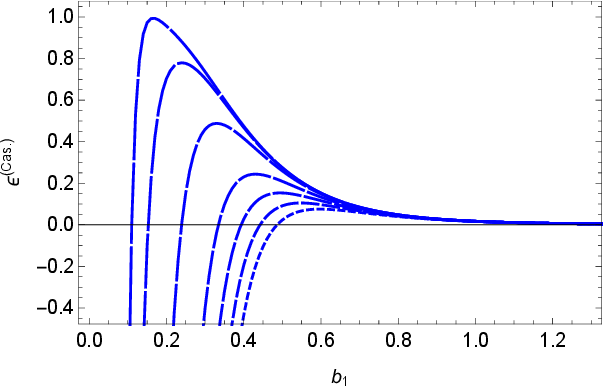}
\includegraphics[width=8cm]{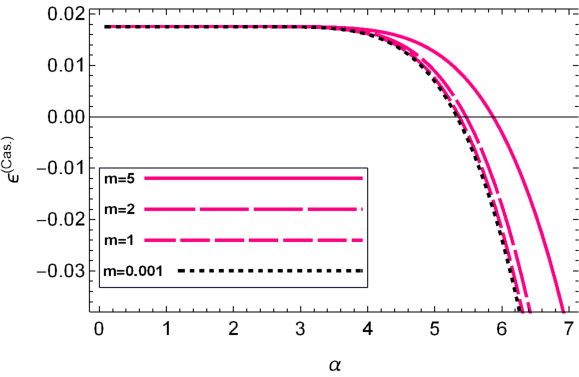}
\caption{\label{PLOT.ROTotalRC}
In the left figure, the total Casimir energy density as a function of radius $b_1$ is plotted for multiple values of the coupling constant \(\alpha = \{3.566, 3.57, 3.6, 3.7, 3.8, 3.9, 4\}\). The smaller dashed lines represent plots for higher values of \(\alpha\). This figure shows that as the coupling constant \(\alpha\) decreases, the transition in sign of the total Casimir energy density occurs at smaller values of the radius \(b_1\). In all plots, the following parameter values were used: \(m_f = 1\), \(\alpha_\psi = 1\), \(m = 1\), and \(\lambda = 0.1\). In the right figure, the total Casimir energy density is plotted as a function of the coupling constant \(\alpha\). This figure displays results for a sequence of boson masses, with \(m = \{5,2, 1, 0.001\}\). The plot indicates that as the boson mass increases, the transition in the sign of the total Casimir energy density occurs at higher values of the coupling constant. The following parameters were used in this plot: \(m_f = 1\), \(\alpha_\psi = 1\), \(b_1 = 1\), and \(\lambda = 0.1\). }
\end{figure}
\begin{figure}[th]\centering\includegraphics[width=10cm]{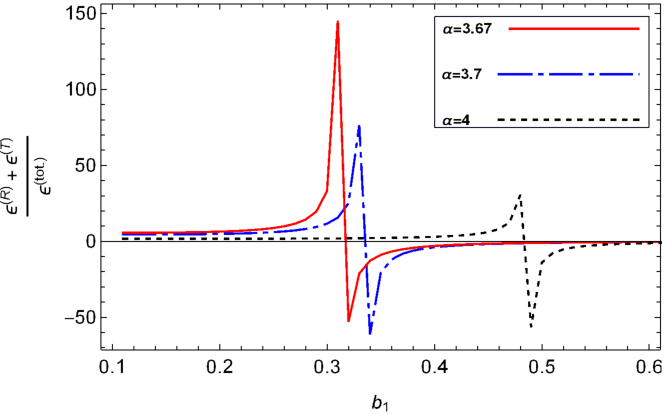}
\caption{\label{PLOT.RORatio}
In this figure, the ratio of the sum of all correction values for the Casimir energy—including radiative and thermal corrections—to the total Casimir energy is plotted as a function of \(b_1\) for three different values of the coupling constant \(\alpha\). The plot reveals that the corrections make a significant contribution to the total Casimir energy, particularly in regions where the sign of the Casimir energy changes. }
\end{figure}

The consistency between the results in Eqs. (\ref{final.EXP.Thermal.MASSIVE.FERMION}) and (\ref{final.EXP.Thermal.MASSLESS.FERMION}) for the thermal correction terms of massive and massless fermionic fields is illustrated in Fig. (\ref{PLOT.ThermalMassCheck}). This figure shows that the thermal correction values for massive and massless fermions are nearly identical when the mass is below $m_f = 0.1$. In the left plot of Fig. (\ref{PLOT.RatioAsMass.AllPlot}), the ratio of the Casimir energy for the massive case to the massless case is plotted as a function of the mass of the field. This plot illustrates the convergence behavior of the Casimir energy between the massive and massless cases. This behavior is depicted simultaneously for the zero-order Casimir energy and its radiative and thermal correction terms in the bosonic case.  Now, we are focusing on a particle spectrum in a bulk and considering the total Casimir energy density within this bulk. The specifics of our chosen particle spectrum in the bulk are detailed in Tab. (\ref{table}). This type of particle spectrum serves as a toy model and is derived from the particle spectrum discussed in Ref. \cite{0905.0328}. In that study, the author introduced a particle spectrum in a bulk and calculated the total Casimir energy density to examine the stabilization of the extra dimension. In our work, we use their toy model to calculate the total Casimir energy density, incorporating a range of corrections. The added correction terms include the radiative correction for massive and massless scalar fields, and the thermal corrections for both bosons and fermions. As shown in Tab. (\ref{table}), the particle spectrum in the bulk comprises a massless boson (such as a graviton), a massless Dirac fermion, a massive Dirac fermion with mass \(m_f\), and massive bosons with equal masses \(m_b\). We adhere to the number of particles (i.e., degrees of freedom) from the toy model documented in Ref. \cite{0905.0328} without altering them. Moreover, a universal coupling constant \(\alpha\) is chosen for both massive and massless scalar fields, while \(\alpha_\psi\) is used for both massive and massless Dirac fields. The main difference between our work and the one reported in Ref. \cite{0905.0328} is the inclusion of radiative and thermal correction terms. Although the effect of radiative correction values is relatively negligible, the combined effect of all corrections, including thermal and radiative corrections for bosons and fermions, can play a significant role in determining the Casimir energy density of the bulk. The sum of all contributions to the Casimir energy density, including the leading order and the correction terms (thermal and radiative) for scalar and Dirac fermion fields, can be expressed as follows:
\begin{eqnarray}\label{SUM.over.CAS.density}
       \mathcal{E}^{\mbox {\tiny (tot.)}}_{\mbox {\tiny Cas.}}=N_b\sum_{\kappa=\{0,1,T\}}\mathcal{E}^{(\kappa)}_{\mbox{\tiny Cas.}}(0,b_1)+\mathcal{N}_b\sum_{\kappa=\{0,1,T\}}\mathcal{E}^{(\kappa)}_{\mbox{\tiny Cas.}}(m_b,b_1)+N_f\sum_{\kappa=\{0,T\}}\mathcal{E}^{(\kappa)}_{\mbox{\tiny Cas.F}}(0,b_1)+N_f\sum_{\kappa=\{0,T\}}\mathcal{E}^{(\kappa)}_{\mbox{\tiny Cas.F}}(m_f,b_1)
\end{eqnarray}
where \(\mathcal{N}_b\) represents the number of bosons, and \(N_f\) represents the number of fermions. It is worth mentioning that the Casimir energy density for massive and massless Dirac fermionic fields coupled with {\ae}ther in compact dimensions has been previously studied in various works. In this study, we utilize the final results from these works as our reference points (for instance, refer to Eqs. (40) and (41) of Ref. \cite{0905.0328}). In the right plot of Fig. (\ref{PLOT.RatioAsMass.AllPlot}), the Casimir energy values of each correction term given in Eq. (\ref{SUM.over.CAS.density}) were plotted as functions of the radius of the compact dimension \( b_1 \). Simultaneously displaying all these correction terms of the Casimir energy side by side in a single diagram can effectively highlight the significance of each term relative to the others. The plot clearly demonstrates that, aside from the radiative correction term, the remaining corrections can significantly influence the total Casimir energy. In the left plot of Fig. (\ref{PLOT.ROTotalRC}), the total Casimir energy density, as defined in Eq. (\ref{SUM.over.CAS.density}), is plotted as a function of the radius \(b_1\). The plot shows that as the coupling constant \(\alpha\) increases with other parameters held constant, the point at which the Casimir energy changes sign occurs at a larger value of \(b_1\). In the right plot of Fig. (\ref{PLOT.ROTotalRC}), it is shown that as the mass of the scalar field decreases relative to the fermionic field, the sign change in the total Casimir energy occurs at a smaller value of the coupling constant. To highlight the significance of the thermal and radiative corrections in the total Casimir energy density, the ratio of the total correction value—including thermal and radiative corrections—to the total Casimir energy density is plotted in Fig. (\ref{PLOT.RORatio}). This plot indicates that these corrections have a significant impact, particularly in regions where the total Casimir energy changes sign.
\section{Conclusions}\label{Sec.Conclusion}
In this study, we explored a five-dimensional spacetime, \((\mathbb{R}^4 \times S^1)\), to determine the radiative correction to the Casimir energy for self-interacting scalar fields, both massive and massless, when coupled with {\ae}ther in a compact fifth space-like dimension. We also calculated the thermal correction to the Casimir energy for this bosonic field, along with the thermal correction for a Dirac fermionic field in the same compact dimension. Using a toy model that represents the degrees of freedom for each boson and fermion in this setup, we derived the total Casimir energy density, accounting for all radiative and thermal corrections. We analyzed the magnitude and sign of this energy density, illustrating the results with figures. Notably, the corrections had a significant impact on the total Casimir energy density, particularly in regions where the sign of the energy density changes. Another important aspect of this study was the use of a distinct renormalization scheme to compute the radiative correction to the Casimir energy, where the counterterms are position-dependent. This approach ensures that the boundary conditions affecting the field are properly accounted in the renormalization program, leading to accurate renormalization of the Lagrangian's bare parameters. Moreover, this type of counterterm aligns with the box subtraction scheme as a regularization technique, providing a self-consistent method for computing the radiative correction to the Casimir energy. By employing this consistent renormalization and regularization approach, we obtained results that satisfy expected physical constraints and deliver coherent answers. This self-consistent method contributes to the reliability and robustness of the computed radiative corrections, offering a comprehensive perspective on the Casimir energy in compact extra-dimensional spaces. In our future works, in addition to studying the Casimir energy of fermionic and bosonic fields, we will consider the effects of Lorentz symmetry breaking on the Casimir energy and its radiative and thermal corrections related to the electromagnetic field in the fifth compact dimension of spacetime. We believe that investigating the total Casimir energy will be more reliable by considering the fluctuations of all possible fields, which can be addressed in future articles.
\appendix
\section{Calculation of Integral Parts}\label{Appendix.Finding Finite Parts}
\setcounter{equation}{0}
The terms \(\mathcal{T}_1\), \(\mathcal{T}_2\), \(\mathcal{T}_3\), and \(\mathcal{T}_4\) in Eq. (\ref{NLO.EXP.3}) are divergent. However, as we will demonstrate, their contributions can be entirely removed by corresponding terms in the vacuum energy of the second system. Additionally, two terms, \(\mathcal{S}_1\) and \(\mathcal{S}_2\), also show divergence. As we will outline below, the infinite parts of these terms are removed through the BSS and cutoff regularization methods, leaving the finite remaining contributions from these terms. To explain this, we begin by:
\begin{eqnarray}\label{T1.}
     \mathcal{T}_1(b_1,\Lambda_1)-\{b_1\to b_2, \Lambda_1\to\Lambda_2\}&=&\frac{1}{2b_1}\left(\frac{b_1}{2\sqrt{1+\alpha^2}}\int_{0}^{\Lambda_1}\frac{\Omega_4 k^3dk}{\sqrt{k^2+m^2}}\right)^2-\{b_1\to b_2, \Lambda_1\to\Lambda_2\}\nonumber\\&=&\frac{b_1\Omega_4^2}{8(1+\alpha^2)}\left[\int_{0}^{\Lambda_1}\frac{k^3dk}{\sqrt{k^2+m^2}}\right]^2-\{b_1\to b_2,\Lambda_1\to\Lambda_2\}\nonumber\\
     &=&\frac{b_1\Omega_4^2}{8(1+\alpha^2)}\mathcal{R}(\Lambda_1)^2-\{b_1\to b_2,\Lambda_1\to\Lambda_2\}.
\end{eqnarray}
By adjusting the cutoffs \(\Lambda_1\) and \(\Lambda_2\) such that \(b_1 \mathcal{R}(\Lambda_1)^2 = b_2 \mathcal{R}(\Lambda_2)^2\), the contributions from \(\mathcal{T}_1\) in Eq. (\ref{NLO.EXP.3}) are eliminated. For the term \(\mathcal{T}_2\), we have:
\begin{eqnarray}\label{T2.}
       \mathcal{T}_2(b_1,\infty)-\{b_1\to b_2\}=\frac{1}{b_1}\left(\int_{0}^{\infty}\frac{4\pi k^2dk}{\sqrt{k^2+m^2}}\right)\frac{b_1}{2\sqrt{1+\alpha^2}}\int_{0}^{\infty}\frac{\Omega_4 \kappa^3d\kappa}{\sqrt{\kappa^2+m^2}}-\{b_1\to b_2\}=0.
\end{eqnarray}
As depicted in Eq. (\ref{T2.}), the term \(\mathcal{T}_2\) is independent of the radii \(b_1\) and \(b_2\). Consequently, the contribution of this term is canceled analytically by a similar term within the second system. A similar cancellation occurs for the term \(\mathcal{T}_3\) from Eq. (\ref{NLO.EXP.3}). Thus, the following holds:
\begin{eqnarray}\label{T3.}
       \mathcal{T}_3(b_1,\infty)-\{b_1\to b_2\}=\frac{1}{4b_1}\frac{b_1}{\sqrt{1+\alpha^2}}\int_{0}^{\infty}dX\left[\int_{0}^{\infty}\frac{4\pi k^2 dk}{\sqrt{k^2+X^2+m^2}}\right]-\{b_1\to b_2\}=0,
\end{eqnarray}
where $X=x\sqrt{1+\alpha^2}/b_1$. For the fourth term, $\mathcal{T}_4$, we get:
\begin{eqnarray}\label{T4.}
      \mathcal{T}_4(b_1;\Lambda_1)-\{b_1\to b_2,\Lambda_1\to\Lambda_2\}&=&\frac{1}{b_1}\int_{0}^{\Lambda_1}\frac{4\pi k^2dk}{\sqrt{k^2+m^2}}-\{b_1\to b_2, \Lambda_1\to\Lambda_2\}\nonumber\\
      &=&\frac{1}{b_1}\mathcal{H}(\Lambda_1)-\frac{1}{b_2}\mathcal{H}(\Lambda_2).
\end{eqnarray}
Cutoff regularization is used to address the infinities in the above expression. By adjusting the cutoffs \(\Lambda_1\) and \(\Lambda_2\) such that \(b_1 \mathcal{H}(\Lambda_2) = b_2 \mathcal{H}(\Lambda_1)\), the divergences resulting from Eq. (\ref{T4.}) are removed. For the term \(\mathcal{S}_1(b_1, \infty)\), we can express it as follows:
\begin{eqnarray}\label{S1}
      \mathcal{S}_1(b_1,\Lambda_1)-\mathcal{S}_1(b_2,\Lambda_2)&=&4\pi m^2
      \left[\int_{0}^{\Lambda_1}\frac{\eta^2d\eta}{\sqrt{\eta^2+1}}\right]
      \mathcal{B}_1(b_1;m)-\{b_1\to b_2,\Lambda_1\to\Lambda_2\}\nonumber\\
      &&\buildrel {\Lambda_1\to\infty}\over\longrightarrow \left[\pi+2\pi\Lambda_1^2-2\pi\ln(2\Lambda_1)+\mathcal{O}(\Lambda_1^{-2})\right]m^2\mathcal{B}_1(b_1;m)-\{b_1\to b_2,\Lambda_1\to\Lambda_2\},
\end{eqnarray}
where $\eta=k/m$. Adjusting the cutoffs with the expression \((\Lambda_1 - \ln(2\Lambda_1))b_2 \mathcal{B}_1(b_1; m) = (\Lambda_2 - \ln(2\Lambda_2)) b_1 \mathcal{B}_1(b_2; m)\) results in the removal of all divergent components of Eq. (\ref{S1}). This adjustment leaves only the finite part, which is $\pi m^2[ \frac{\mathcal{B}_1(b_1; m)}{b_1} - \frac{\mathcal{B}_1(b_2; m)}{b_2}]$. We apply a similar cancellation process to the following term
\begin{eqnarray}\label{S2.}
      \mathcal{S}_2(b_1,\Lambda_1)-\mathcal{S}_2(b_2,\Lambda_2)&=&\mathcal{B}_1(b_1;m)\frac{m^3 \Omega_4}{2\sqrt{1+\alpha^2}}\int_{0}^{\Lambda_1}\frac{\eta^3d\eta}{\sqrt{\eta^2+1}}-\{b_1\to b_2,\Lambda_1\to\Lambda_2\}\nonumber\\&=&\frac{ m^3 \pi^2}{3\sqrt{1+\alpha^2}} \left(\Lambda ^2\sqrt{\Lambda ^2+1}-2\sqrt{\Lambda^2+1}+2\right)\mathcal{B}_1(b_1,m)\nonumber\\&& \buildrel {\Lambda_1\to\infty}\over\longrightarrow \frac{m^3\pi^2}{\sqrt{1+\alpha^2}}\left[\frac{2}{3}-\frac{\Lambda}{2}+\frac{\Lambda^3}{3}+\mathcal{O}(\Lambda_1^{-1})\right]\mathcal{B}_1(b_1;m),
\end{eqnarray}
where $\eta=k/m$. By adjusting the cutoffs according to the relationship \(\mathcal{B}_1(b_1, m) \left(-\frac{\Lambda_1}{2} + \frac{\Lambda_1^3}{3}\right) = \mathcal{B}_1(b_2, m) \left(-\frac{\Lambda_2}{2} + \frac{\Lambda_2^3}{3}\right)\), all infinite components from the previous equation are eliminated. The only finite contribution that remains is \(\frac{2 m^3 \pi^2}{3\sqrt{1+\alpha^2}} \mathcal{B}_1(b_1; m) - \{b_1 \rightarrow b_2\}\). Combining all results from this appendix, Eq. (\ref{NLO.EXP.3}) is transformed into Eq. (\ref{Eq.After.Appen.}).


\begin{acknowledgments}
The author would like to thank the research office of Semnan Branch, Islamic Azad University for the financial support.
\end{acknowledgments}

\begin{thebibliography}{99}
\bibitem{h.b.g.}
         H. B. G. Casimir, Proc. Kon. Nederl. Akad. Wet. {\bf 51} (1948) 793.
\bibitem{spaarnay}
         M. J. Sparnaay, Physica {\bf 24} (1958) 751.
\bibitem{other.works.1}
         K.A. Milton, \emph{The Casimir Effect: Physical Manifestations of Zero-Point Energy} (Singapore: World Scientific) 2001.
\bibitem{Lamerux}
         S. K. Lamoreaux, \emph{Demonstration of the Casimir Force in the $0.6$ to $6\mu m$ Range}, Phys. Rev. Lett., {\bf78} (1997) 5–8.
\bibitem{Physica.scr.1}
         V. Abhignan, \emph{Casimir energy of N magnetodielectric $\delta$-function plates}, Phys. Scr. {\bf98} (2023) 105018.
\bibitem{Physica.scr.2}
         V. V. Dodonov, \emph{Current status of the dynamical Casimir effect}, Phys. Scr. {\bf82} (2010) 038105.
\bibitem{bio.Casimir.1}
         P. Pawlowski and P. Zielenkiewicz, \emph{The quantum casimir effect may be a universal force organizing the bilayer structure of the cell membrane}, The Journal of membrane biology \textbf{246} (2013) 383-389.
\bibitem{bio.Casimir.2}
        A. Gambassi, \emph{The Casimir effect: From quantum to critical fluctuations}, J. Phys. Conf. Ser. \textbf{161} (2009) 012037, [arXiv:0812.0935].
\bibitem{bio.Casimir.3}
         B. B. Machta, S. L. Veatch, and J. P. Sethna, \emph{Critical casimir forces in cellular membranes}, Phys. Rev. Lett. \textbf{109} (2012) 138101.
\bibitem{other.works.2}
         M. Bordag, G.L. Klimchitskaya, U. Mohideen and V.M. Mostepanenko, \emph{Advances in the Casimir Effect} (New York: Oxford University Press) 2009.
\bibitem{RC.1}
         L.H. Ford, \emph{Casimir effect for a self-interacting scalar field}, Proc. R. Soc. A {\bf368} (1979) 305
\bibitem{JPhysG.Man}
       M.A. Valuyan, \emph{The Dirichlet Casimir energy for $\phi^4$ theory in a rectangular waveguide}, J. Phys. G: Nucl. Part. Phys. {\bf45} (2018) 095006 (23pp)
\bibitem{Bordag.1}
         M. Bordag, D. Robaschik and E. Wieczorek, \emph{Quantum field theoretic treatment of the Casimir effect}, Ann. Phys. {\bf165} (1985) 192.
\bibitem{Bordag.2}
       M. Bordag and K. Scharnhorst, \emph{O($\alpha$) radiative correction to the Casimir energy for penetrable mirrors}, Phys. Rev. Lett. {\bf81} (1998) 3815.
\bibitem{Bordag.3}
       M. Bordag and J. Lindig, \emph{Radiative correction to the Casimir force on a sphere}, Phys. Rev. D {\bf58} (1998) 045003.


\bibitem{thermal.Brevik.}
         Iver Brevik, S. A. Ellingsen and K. A. Milton, \emph{Thermal corrections to the Casimir effect}, New J. Phys. {\bf8} (2006) 236.
\bibitem{all.method.ZO}
         M. Sasanpour, C. Ajilyan and Siamak S. Gousheh, \emph{Casimir free energy for massive fermions: a comparative study of various approaches}, J. Phys. A: Math. Theor. {\bf55} (2022) 125401.


\bibitem{methods.ZetaF}
        Andrea Erdas and Kevin P. Seltzer, \emph{Finite temperature Casimir effect for charged massless scalars in a magnetic field}, Phys. Rev. D \textbf{88} (2013) 105007.
\bibitem{Heat.kernel}
        V.V. Nesterenko, G. Lambiaseb, G. Scarpetta, \emph{Calculation of the Casimir energy at zero and finite temperature: some recent results}, Riv. Nuovo Cim. \textbf{27} (2004) 1-74; [arXiv: hep-th/0503100].
\bibitem{methods.BSS}
       M.A. Valuyan, \emph{The Casimir energy for scalar field with mixed boundary condition}, Int. J. Geo. Meth. Mod. Phys. {\bf15} (2018) 1850172.
\bibitem{cognola}
       G. Cognola, E. Elizalde and K. Kirsten, \emph{Casimir energies for spherically symmetric cavities}, J. Phys. A: Math. Gen. {\bf34} (2001) 7311.
\bibitem{cruz.Fermion}
D. R. da Silva, M. B. Cruz and E. R. Bezerra de Mello, \emph{Fermionic Casimir effect in Horava–Lifshitz theories}, Int. J. Mod. Phys. A {\bf34} (2019) 1950107.
\bibitem{cruz.1}
        M. B. Cruz, E. R. Bezerra de Mello and A. Yu. Petrov, \emph{Casimir effects in Lorentz-violating scalar field theory}, Phys. Rev. D \textbf{96} (2017) 045019.
\bibitem{cruz.2}
        M. B. Cruz, E. R. Bezerra de Mello and A. Yu. Petrov, \emph{Fermionic Casimir effect in a field theory model with Lorentz symmetry violation}, Phys. Rev. D \textbf{99} (2019) 085012.
\bibitem{Lv.1}
         Hiroki Matsui, Yutaka Sakamura, \emph{Pauli–Villars Regularization of Kaluza–Klein Casimir Energy with Lorentz Symmetry}, Prog. Theo. and Exp. Phys., Volume \textbf{2024} (2024) 043B06.
\bibitem{Aether.EM.}
       K.E.L. de Farias, M. A. Anacleto, E. Passos, Iver Brevik, Herondy Mota, and Jo\~{a}o R.L. Santos, \emph{{\AE}ther-electromagnetic theory and the Casimir effect}, Phys. Rev. D {\bf109} (2024) 125010.
\bibitem{Mojavezi.}
       A. Mojavezi, R. Moazzemi, M. E. Zomorrodian, \emph{NLO radiative correction to the Casimir energy in Lorentz-violating scalar field theory}, Nucl. Phys. B {\bf941} (2019) 145-157.
\bibitem{9ta12.0905.0328.1}
        S.M. Carroll and E.A. Lim, \emph{Lorentz-Violating Vector Fields Slow the Universe Down}, Phys. Rev. D {\bf70} (2004) 123525; [arXiv:hep-th/0407149].
\bibitem{9ta12.0905.0328.2}
        B. Li, D.F. Mota and J. D. Barrow, \emph{Detecting a Lorentz-Violating Field in Cosmology}, Phys. Rev. D {\bf77} (2008) 024032; [arXiv:0709.4581].
\bibitem{9ta12.0905.0328.3}
        R.K. Obousy and G. Cleaver,\emph{ Radius Destabilization in Five Dimensional Orbifolds Due to an Enhanced Casimir Effect}, Modern Physics Letters A {\bf24} (2009) 1495-1506.
\bibitem{9ta12.0905.0328.4}
        R.K. Obousy and G. Cleaver, \emph{Casimir Energy and Brane Stability}, J. Geo. and Phys. {\bf61} (2011) 577-588.
\bibitem{Greene.}
       B.R. Greene and J. Levin, Dark Energy and Stabilization of Extra Dimensions, JHEP {\bf11} (2007) 096; [arXiv:0707.1062].
\bibitem{Ref.1}
       P. Burikham, A. Chatrabhuti, P. Patcharamaneepakorn and K. Pimsamarn, \emph{Dark energy and moduli stabilization of extra dimensions in $M^{1+3}\times T^2$ spacetime}, JHEP {\bf07} (2008) 013 [arXiv:0802.3564].
\bibitem{Ref.2}
       N. Arkani-Hamed, S. Dimopoulos, and G. Dvali, \emph{The Hierarchy Problem and New Dimensions at a Millimeter}, Phys. Lett. B {\bf429} (1998) 263; [arXiv:hep-ph/9803315].
\bibitem{Ref.3}
       I. Antoniadis, N. Arkani-Hamed, S. Dimopoulos, and G. Dvali, \emph{New Dimensions at a Millimeter to a Fermi and Superstrings at a TeV}, Phys. Lett. B {\bf436} (1998) 257; [arXiv:hep-ph/9804398].
\bibitem{0905.0328}
       A. Chatrabhuti, P. Patcharamaneepakorn, and P. Wongjun, \emph{Aether Field, Casimir Energy and Stabilization of The Extra Dimension}, JHEP {\bf08} (2009) 019.
\bibitem{free.counterterm.1}
        N. Graham, R. L. Jaffe, V. Khemani, M. Quandt, O. Schroeder and H. Weigel, \emph{The Dirichlet Casimir problem}, Nucl. Phys. B \textbf{677} (2004) 379.
\bibitem{free.counterterm.2}
        M. I. Caicedo and N. F. Svaiter, \emph{Effective Lagrangians for scalar fields and finite size effects in field theory}, J. Math. Phys. \textbf{45} (2004) 179.
\bibitem{pos.dep.counterterm.}
        A. Mojavezi, R. Moazzemi and M. E. Zomorrodian, \emph{Coordinate space representation for renormalization of quantum electrodynamics}, Eur. Phys. J. Plus {\bf136} (2021) 200.
\bibitem{koli.1}
E. Farhi, N. Graham, R.L. Jaffe, H. Weigel, \emph{Heavy Fermion Stabilization of Solitons in 1+1 Dimensions}, Nucl. Phys. B \textbf{585} (2000) 443.
\bibitem{koli.2}
            A.A. Izquierdo, W.G. Fuertes, M.A. Gonzâlez Leôn, \emph{One-loop corrections to classical masses of kink families},  J.M. Guilarte, Nucl. Phys. B \textbf{681} (2004) 163.
\bibitem{koli.3}
             A.S. Goldhaber, A. Litvintsev, P. van Nieuwenhuizen, \emph{Mode regularization of the supersymmetric sphaleron and kink: Zero modes and discrete gauge symmetry} Phys. Rev. D \textbf{64} (2001) 045013.
\bibitem{fosco.}
          C.D. Fosco, N.F. Svaiter, \emph{Finite Size Effects in the Anisotropic $(\lambda/4!)(\phi^4_1 + \phi^4_2)_d$ Model},  J. Math. Phys. \textbf{42} (2001) 5185.
\bibitem{freevapositioncounterterm}
           L.C. de Albuquerque, \emph{Casimir pressure at two loops and soft boundaries at finite temperature}, Phys. Rev. D \textbf{55} (1997) 7754.
\bibitem{2D.Man}
           S.S. Gousheh, R. Moazzemi, M.A. Valuyan, \emph{Radiative correction to the Dirichlet Casimir energy for $\lambda\phi^4$ theory in two spatial dimensions}, Phys. Lett. B \textbf{681} (2009) 477–483.
\bibitem{APSF.Ref.1}
             P. Henrici, \emph{Applied and Computational Complex Analysis}, vol. 1, Wiley, New York, 1984.
\bibitem{APSF.Ref.2}
            E.T. Whittaker, G.N. Watson, \emph{A Course of Modern Analysis}, Cambridge Univ. Press, 1958.
\bibitem{APSF.Ref.3}
            A. A. Saharian, \emph{The generalized Abel-Plana formula. Applications to Bessel functions and Casimir effect}, arXiv:hep-th/0002239v1.
\bibitem{Fermion.Lagrangian}
            S.M. Carroll and H. Tam, \emph{Aether compactification}, Phys. Rev. D {\bf78} (2008) 044047; [arXiv:0802.0521].





































\end{thebibliography}

\end{document}